\documentclass{aa}

\usepackage{txfonts}
\usepackage{natbib}
\bibpunct{(}{)}{;}{a}{}{,}
\usepackage{graphicx}

\def\be{\begin{equation}}
\def\ee{\end{equation}}
\def\bea{\begin{eqnarray}}
\def\eea{\end{eqnarray}}

\def\bB{\vec{B}}
\def\bb{\vec{b}}
\def\bu{\vec{u}}
\def\br{\vec{r}}
\def\bJ{\vec{J}}
\def\bj{\vec{j}}
\def\bA{\vec{A}}
\def\ba{\vec{a}}

\def\hB{\vec{\hat{B}}}
\def\hb{\vec{\hat{b}}}
\def\hJ{\vec{\hat{J}}}
\def\hj{\vec{\hat{\jmath}}}
\def\ha{\vec{\hat{a}}}
\def\hr{\vec{\hat{r}}}

\def\balpha{\vec{\alpha}}
\def\bbeta{\vec{\beta}}
\def\bn{\vec{\nabla}}
\def\bsigma{\vec{\sigma}}

\def\t{\!\times\!}
\def\cd{\!\cdot\!}

\def\d{\mathrm{d}}

\def\lam{\lambda}
\def\al{\alpha}
\def\Om{\Omega}
\def\p{\partial}
\def\th{\vartheta}
\def\phi{\varphi}

\begin{document}

\title{An efficient method for computing the eigenfunctions\\of the dynamo equation}

\author{M. Schrinner\inst{1,2} \and
D. Schmitt\inst{1}\fnmsep\thanks{Corresponding author, e-mail: schmitt@mps.mpg.de}
\and J. Jiang\inst{1} \and P. Hoyng\inst{3}}

\institute{Max-Planck-Institut f\"ur Sonnensystemforschung, Max-Planck-Str. 2, 37191 Katlenburg-Lindau, Germany \and
MAG (ENS/IPGP), LRA, Ecole Normale Sup\'erieure, 24 rue Lhomond, 75252 Paris Cedex 05, France \and
SRON Netherlands Institute for Space Research, Sorbonnelaan 2, 3584 CA Utrecht, The Netherlands}

\titlerunning{An efficient method for computing the eigenfunctions of the dynamo equation}
\authorrunning{M. Schrinner et al.}

\date{Received 19 November 2009 / Accepted 19 May 2010}

\abstract{}{We present an elegant method of determining the eigensolutions of
the induction and the dynamo equation in a fluid embedded in a vacuum.}{The
magnetic field is expanded in a complete set of functions. The new method is
based on the biorthogonality of the adjoint electric current and the vector
potential with an inner product defined by a volume integral over the fluid
domain. The advantage of this method is that the velocity and the dynamo
coefficients of the induction and the dynamo equation do not have to be
differentiated and thus even numerically determined tabulated values of the
coefficients produce reasonable results.}{We provide test calculations and
compare with published results obtained by the classical treatment based on the
biorthogonality of the magnetic field and its adjoint. We especially consider
dynamos with mean-field coefficients determined from direct numerical
simulations of the geodynamo and compare with initial value calculations and
the full MHD simulations.}{}


\maketitle

\section{Introduction}

The generation and evolution of magnetic fields in cosmic bodies like the
planets and stars is generally thought to be governed by induction processes
due to motions in their electrically conducting fluid interior. The magnetic
field $\bB$ is described by the induction equation
\be \frac{\p\bB}{\p t} = \bn \t D\bB \label{ie} \ee
where
\be D\bB = \bu\t\bB - \eta\bn\t\bB \;. \label{io} \ee
Here $\bu$ represents the velocity and $\eta$ the magnetic diffusivity.

In the framework of mean-field theory \citep[e.g.][]{m78,kr80}, $\bu$ and $\bB$
are considered as mean, e.g. ensemble averaged, quantities, whereas the action
of the small-scale turbulent flow on the mean magnetic field is parametrised by
the so-called dynamo coefficients, $\balpha$ and $\bbeta$. They are, in
general, tensors of second and third rank, respectively. We use the following
compact notation of the mean field coefficients, which include the so-called
$\gamma$, $\delta$, and $\kappa$-effects, see e.g. \citet{r80}. Then, the
operator $D$ reads
\be D\bB = \bu\t\bB + \balpha\cd\bB - \bbeta:(\bn\bB) - \eta\bn\t\bB
\label{do} \ee
instead of (\ref{io}), and acts on the mean magnetic field. Except for the
additional $\balpha$ and $\bbeta$ terms in the $D$ operator, the induction and
the dynamo equation are formally equivalent. Thus, the new method presented
here equally applies to both.

The dynamo region is located in a flow domain $V$ with exterior vacuum $E$. In
this work we assume $V$ to be either a sphere or a spherical shell. The
magnetic field $\bB$ is continuous through the boundary $\p V$ and potential in
$E$.

In kinematic dynamo theory, all coefficients ($\bu$, $\balpha$, $\bbeta$, and
$\eta$) are assumed given and independent of the magnetic field. Thus the
dynamo equation is linear in the magnetic field and can be solved by
considering an eigenvalue problem
\be \lam\bB = \bn \t D\bB \label{dep} \ee
with eigenvalues $\lam$ describing the time evolution proportional to
$\exp(\lam t)$ of the magnetic field $\bB$.

Many studies have been made of the eigenvalues of the dynamo operator for
various celestial bodies and with many forms of the dynamo coefficients
\citep[e.g.][]{bg54,r60,sk69,ds71,r72,rs72,g73,kr75,ss89,dj89,dgs93,g00,sz01,
lj04,lj05,jw06,jw07}. Often the coefficients are approximated by simple
analytical functions of position, and their tensorial character is disregarded.
Recently, the test-field method, developed by \citet{srsrc05,srsrc07}
\citep[see also][]{osb01,osbr02}, allows one to determine all tensorial
components of $\balpha$ and $\bbeta$ directly from self-consistent numerical
simulations \citep{b08,k09}. These coefficients are sometimes strongly varying
functions of position. This may introduce large errors because the dynamo
operator $\bn\t D$ involves differentiation of the dynamo coefficients, and
these are only available as numerically determined tabulated values.

In this paper we present a new method that does not require differentiation, so
it is also applicable to numerically determined dynamo coefficients. The method
is based on the biorthogonality of the electric current and the vector
potential with an inner product defined by a volume integral over the fluid.
This property has already been noted by \citet{rb87}, \citet{h88},
\citet{frs93}, \citet{hs95}, and \citet{rraf02}.

The method is described in detail in Sect.~\ref{sec_ep}. Extensive test
calculations have been performed and compared with published results by other
eigenvalue methods. Some of these tests are presented in Sect.~\ref{sec_tr}. In
Sect.~\ref{sec_gd} we apply the new method to eigenmodes of the dynamo operator
with coefficients obtained from geodynamo models. Our conclusion are drawn in
Sect.~\ref{sec_concl}.

\section{Eigenvalue problem} \label{sec_ep}

We expand the field $\bB$ of the dynamo in a complete set of functions
$\bb_i(\br)$:
\be \bB = \sum_i e_i\bb_i(\br) \equiv e_i\bb_i \;. \label{ex} \ee
Here and in the following we make use of the summation convention for two
identical indices. The expansion functions are often eigenfunctions of some
differential operator. Since this operator is, in general, not self-adjoint,
the functions are not orthogonal. This problem is handled by using the adjoint
set $\hb_k(\br)$, with the following inner product
\be (\hb_k,\bb_i)_X \equiv \int_X \hb_k\cd\bb_i \, \d^3\br = \delta_{ki}
\;.\label{oip} \ee
The integration volume $X$ can be either the whole space $V+E$ or the fluid
domain $V$ alone. The base functions $\hb_k(\br)$ and $\bb_i(\br)$ constitute a
biorthogonal set. For a given set of functions $\bb_i$, the adjoint set $\hb_k$
depends on the choice of the integration domain $X$, so in principle we have
two different sets $\hb_k$, one for $X=V$ and one for $X=V+E$.

Later we adopt the free magnetic decay modes, for which the base functions
$\bb_i$ and their adjoints $\hb_k$ are known. But at this point there is no
need to specify which set $\bb_i$ we actually use.

\subsection{Biorthogonal sets} \label{bos}

Starting from a set $\hb_k$ and $\bb_i$ that is biorthogonal on $V+E$, a very
useful biorthogonal set on $V$ is provided by the associated electric current
$\hj=\bn\times\hb$ and the vector potential $\ba$ where $\bn\times\ba=\bb$,
with the inner product
\be (\hb_k,\bb_i)_{V+E} = (\hj_k,\ba_i)_V = (\ha_k,\bj_i)_V =\delta_{ki}
\label{nip} \;. \ee
Here we have absorbed a factor of $4\pi/c$ in the definition of the current
$\bj$. The relation (\ref{nip}) is derived with the help of the vector identity
$\bn\cdot(\ba_i\times\hb_k)=\hb_k\cdot(\bn\times\ba_i)-\ba_i\cdot(\bn\times\hb_k)
=\hb_k\cdot\bb_i-\ba_i\cdot\hj_k$ and a volume integration over $V+E$; $\ba$
and $\bb$ go fast enough to zero at infinity. Surface integrals vanish because
the field and the vector potential are continuous through $\p V$. Volume
integrals containing currents are restricted to $V$ since $\bj=0$ in $E$.
The inner product (\ref{nip}) is invariant under a gauge transformation
$\ba\to\ba+\bn\psi$ because $\int_V\hj\cdot\bn\psi\,\d^3\br=
\int_V\bn\cdot(\psi\hj)\,\d^3\br=\int_{\p V}\psi\hj\cdot\d^2\bsigma=0$, as
currents and their adjoints run parallel to the boundary.

Electric currents and vector potentials thus form a biorthogonal set on $V$.
This is essential for the new eigenvalue method presented in Sect.
\ref{nem}.

\subsection{Classical eigenvalue method} \label{cem}

Inserting the expansion (\ref{ex}) in the dynamo eigenvalue equation
(\ref{dep}) yields
\be \lam e_i\bb_i = \bn\t(De_i\bb_i) \;. \label{ce1} \ee
Subsequently, we take the inner product (\ref{oip}) based on $V$ with the
adjoint magnetic field. This leads to
\be \lam e_k = M_{ki}e_i \;\;\mathrm{with}\;\; M_{ki}=(\hb_k,\bn\t D\bb_i)_V
\;. \label{ce2} \ee
A partial integration to shift the curl from the second to the first term, as
done in (\ref{nip}) above and used in the new method below, is not possible
because the surface term $\int_{\p V}(D\bb_i\t\hb_k)\cdot\d^2\bsigma$ need not
vanish here.

We mention as an aside that the magnetic field is often decomposed in its
poloidal and toroidal components (see Appendix \ref{dm}) after which the dynamo
equation is formulated in terms of the defining scalars $P$ and $T$. If the
dynamo coefficients possess certain symmetry properties, the solutions can be
split into two independent subsets, describing magnetic fields symmetric and
antisymmetric with respect to the equator.

\subsection{New eigenvalue method} \label{nem}

We start again with (\ref{ce1}), which we uncurl to obtain
\be \lam e_i\ba_i = De_i\bb_i +\bn\psi \;. \label{ne1} \ee
Taking now the inner product (\ref{nip}) with the adjoint current results in
\be \lam e_k = N_{ki}e_i \;\;\mathrm{with}\;\; N_{ki}=(\hj_k,D\bb_i)_V \;.
\label{ne2} \ee
The gradient term drops out as discussed in Sect.~\ref{bos} above. The
corresponding adjoint functions $\hb_k$ here are different from those in
Sect.~\ref{cem} as they pertain to a different inner product.

The matrices $M_{ki}$ and $N_{ki}$ have the same eigenvalues $\lam$. The
advantage of the new method using $N_{ki}$ in (\ref{ne2}) instead of $M_{ki}$
in (\ref{ce2}) is that no differentiation of the operator $D$ is required, so
even numerically computed or tabulated values of $\bu$, $\balpha$, and $\bbeta$
produce accurate results.

\subsection{Choice of $\bb_i$ and numerical handling of (\ref{ne2})} \label{num}

For the set of base functions, we adopt the free magnetic decay modes whose
magnetic fields $\bb_i$ are known analytically in $V+E$ in terms of the
defining scalars $P$ and $T$ as described in Appendix \ref{dm}. The decay modes
are continuous through $\p V$ and potential in $E$, so they satisfy the
boundary conditions. They are characterised by three numbers, the radial order
$n$, the latitudinal degree $l$, and the azimuthal order $m$.

Another advantage of the decay modes is that they are self-adjoint on $V+E$ so
that the adjoint functions are the complex conjugates $\hb_k=\bb^*_k$ and
likewise $\hj_k=\bj^*_k$. Normalisation on $V+E$, i.e.
$(\bb_k^*,\bb_i)_{V+E}=(\bj_k^*,\ba_i)_V=\delta_{ki}$, is thus readily
achieved, see Appendix \ref{dm}.

The computation of the matrix elements $N_{ki}$ is now straightforward. Once we
know the matrix elements, the eigenvalue problem (\ref{ne2}) is solved
numerically using LAPACK routines (http://www.netlib.org/lapack), and we obtain
the eigenvalues $\lam_k$ and eigenvectors $\{e_{ki}\}$, such that
\be\bB_k=e_{ki}\bb_i\ee is eigenfunction of $\bn\times D$ with eigenvalue
$\lam_k$. Each mode $k$ contains, in general, a mixture of $n$, $l$, and $m$
values.

\begin{table}
\caption{Eigenvalues of the fundamental dipolar mode and the fifth and tenth
overtones of the $\al^2$-sphere.} \label{tab_al2}
\begin{center}
\begin{tabular}{cccc}
\hline\hline
\rule[0.0mm]{0mm}{2.2mm} $n_\mathrm{max}$ & $\lam_0$ & $\lam_5$ & $\lam_{10}$ \\
\hline
 4 & $-$0.0241 & $-$116.936 \\
 8 & $-$0.0034 & $-$115.110 & $-$349.489 \\
12 & $-$0.0010 & $-$115.058 & $-$349.097 \\
16 & $-$0.0004 & $-$115.047 & $-$349.050 \\
20 & $-$0.0002 & $-$115.043 & $-$349.036 \\
24 & $-$0.0001 & $-$115.041 & $-$349.031 \\
28 & $-$0.0001 & $-$115.040 & $-$349.028 \\
32 & $-$0.0000 & $-$115.040 & $-$349.026 \\
36 & $-$0.0000 & $-$115.039 & $-$349.026 \\
\hline
HvG93 & 0.0000 & $-$115.04 & $-$349.02 \\
\hline
\end{tabular}
\end{center}
\end{table}

In the following we consider only velocities and dynamo coefficients that are
independent of azimuth $\phi$, but this is not a necessary constraint. Thus
each value of $m$ can be treated separately. Although we present only results
for $m=0$ here, we have tested and applied other values of $m$ as well. We
employ the robust Gauss-Legendre quadrature in $r$ and $\cos\th$ to compute the
matrix elements since the basis functions are heavily oscillatory in $r$ for
high values of $n$ and in $\theta$ for high degree $l$. For the Gauss-Legendre
integration we used 66 quadrature points here in the radial and 80 in the
latitudinal direction, respectively. In general, this depends of course on the
required resolution.

\subsection{Adjoint eigenfunctions} \label{ae}

We now show how one may construct the adjoint set of eigenfunctions $\hB_p$ of
a set of eigenfunctions $\bB_i$ of the dynamo operator $\bn\t D$. Although
these adjoints are not needed in the present paper, they appear in
applications. For example, let $\bB$ be the actual magnetic field of the
dynamo, then it is often advantageous to expand $\bB$ in dynamo eigenfunctions,
i.e., $\bB(\br,t)=\sum_i c_i(t)\bB_i(\br)$. To find the coefficients $c_i(t)$
we use the adjoint set, to find $c_i=(\hB_i,\bB)_{V+E}=(\hJ_i,\bA)_V$.

This illustrates that we need the adjoints $\hB_p$, and these may be
constructed as follows. Let $\bB_k=e_{ki}\bb_i$ be the representation of
$\bB_k$ in terms of the self-adjoint magnetic decay modes as above. Then we
write $\hB_p=f^*_{pi}\bb^*_i$ and $\hJ_p=f^*_{pi}\bj^*_i$, and we require
\be \delta_{pk}=(\hJ_p,\bA_k)_V=(f^*_{pi}\bj^*_i,e_{kj}\ba_j)_V=
f^*_{pi}e_{kj}\delta_{ij}=f^*_{pi}e_{ki}\,. \ee
Therefore $\mathrm{f}^\dagger=\mathrm{e}^{-1}$ in matrix notation, and we find
a unique biorthogonal set. Here, $\dagger$ indicates the Hermitean adjoint,
$\mathrm{f}^\dagger=(\mathrm{f}^T)^*$, where $T$ indicates the transposed and
$*$ complex conjugation.

Three important messages follow from this construction: (i) the adjoint of
$\bn\times\bB$ is $\bn\times\hB$, that is, the adjoint operation commutes with
$\bn$; (ii) to obtain the adjoint eigenfunctions, it is not necessary to know
the explicit form of the adjoint dynamo operator $\bn\times\hat{D}$; and (iii)
the eigenfunctions and their adjoints have the same boundary conditions because
they are a linear combination of the decay modes and their complex conjugates,
respectively.

\begin{table}
\caption{Eigenvalues of the fundamental mode for the $\vec{t_1s_2}$ flow for
two magnetic Reynolds numbers $R_m$.} \label{tab_sf}
\begin{center}
\begin{tabular}{ccc}
\hline\hline
$n_\mathrm{max}=l_\mathrm{max}$ & $\lam_0 \; (R_m=10)$ & $\lam_0 \; (R_m=100)$ \\
\hline
 4 & $-$8.02911 & $-$5.88075 \\
 8 & $-$8.02625 & $-$6.84495 \\
12 & $-$8.02626 & $-$6.92613 \\
16 & $-$8.02627 & $-$6.92859 \\
20 & $-$8.02627 & $-$6.92866 \\
24 & $-$8.02627 & $-$6.92869 \\
\hline
LJ05 & $-$8.01600 & $-$6.92885 \\
\hline
\end{tabular}
\end{center}
\end{table}

\section{Test results} \label{sec_tr}

\subsection{$\al^2$-sphere}

We first consider the so-called $\al^2$-sphere of unit radius $r_0=1$,
represented by $\bu=\bbeta=\vec{0}$, $\al_{ij}=R_\al\delta_{ij}$, and $\eta=1$,
which also can be treated analytically \citep[Chap.~14]{kr80}. The eigenvalues
are independent of azimuth $m$, and the eigenfunctions decouple in latitudinal
quantum number $l$. For $R_\al\neq0$, the modes couple in radial number $n$, as
they do between the poloidal and toroidal components.

For $R_\al=4.493409458$,\footnote{The numerical value of $R_\al$ is equal to
the first zero of the spherical Bessel function $j_1$.} the first mode is a
stationary dipole, while the overtones decay with the rates given by
\citet{hg93} (HvG93). We successfully reproduced the fundamental mode and the
overtones. In Table~\ref{tab_al2} we consider the convergence in the
eigenvalues as a function of the maximum radial number $n_\mathrm{max}$ for
some dipolar $(l=1)$ modes. Higher $l$ modes behave similarly. We also
reproduced the eigenfunction plots as provided by \citet{kr80}.

\begin{table*}
\caption{Eigenvalues of the first and fifth modes of the $\al^2\Om$-dynamo (see
Sect.~\ref{sec_aod}).} \label{tab_jie}
\begin{center}
\begin{tabular}{cc|cc|cc}
\hline\hline
& & \multicolumn{2}{c}{New method} & \multicolumn{2}{c}{Classical method} \\
\hline
$n_\mathrm{max}$ & $l_\mathrm{max}$ & $\lam_0 \; (S)$ & $\lam_4 \; (A)$ &
$\lam_0 \; (S)$ & $\lam_4 \; (A)$ \\
\hline
 8 &  8 & (16.572\,,\,0.0) & ($-$25.767\,,\,$\pm$45.538) & (16.577\,,\,0.0) & ($-$25.763\,,\,$\pm$45.529) \\
12 & 12 & (16.053\,,\,0.0) & ($-$33.126\,,\,$\pm$39.444) & (16.053\,,\,0.0) & ($-$33.125\,,\,$\pm$39.443) \\
16 & 16 & (16.052\,,\,0.0) & ($-$33.201\,,\,$\pm$40.168) & (16.052\,,\,0.0) & ($-$33.201\,,\,$\pm$40.168) \\
20 & 20 & (16.053\,,\,0.0) & ($-$33.200\,,\,$\pm$40.157) & (16.053\,,\,0.0) & ($-$33.200\,,\,$\pm$40.157) \\
\hline
 8 & 20 & (16.052\,,\,0.0) & ($-$33.203\,,\,$\pm$40.158) & (16.057\,,\,0.0) & ($-$33.197\,,\,$\pm$40.163) \\
12 & 20 & (16.052\,,\,0.0) & ($-$33.200\,,\,$\pm$40.157) & (16.052\,,\,0.0) & ($-$33.199\,,\,$\pm$40.157) \\
\hline
\end{tabular}
\end{center}
\end{table*}

\subsection{Spherical flows}

As a next test, we apply the spherical stationary $\vec{t_1s_2}$ (MDJ) flow of
\citet{lj04} which is given by \bea
\bu &=& u_0K^{-1}\bn\t\left(r^2(1-r^2)P_1^0(\cos\th)\hr\right) \nonumber \\
&&
\null+u_0K^{-1}\epsilon\,\bn\t\bn\t\left(r^3(1-r^2)^2P_2^0(\cos\th)\hr\right)
\eea with $K^{-1}=\sqrt{9009/572}$ and $\epsilon=0.5\sqrt{143/1008}$ such that
the rms poloidal to toroidal energy ratio is 0.5, and the flow has an rms value
of $u_0$. $P_l^m$ and $\hr$ are defined in Appendix~\ref{dm}.

Like \citet{lj05} we consider the axisymmetric $(m=0)$ and equatorially
antisymmetric magnetic field solution for a unit sphere $(r_0=1)$ embedded in a
vacuum. Table~\ref{tab_sf} shows the convergence in the eigenvalue with the
largest real part as a function of truncation $n_\mathrm{max}$ and
$l_\mathrm{max}$ for two magnetic Reynolds numbers $R_m=u_0r_0/\eta=10$ and
$R_m=100$, together with the converged values given by \citet{lj05} (LJ05).
There is a difference of about one permille between their value and ours for
$R_m=10$.

\subsection{$\al^2$ and $\al^2\Om$-dynamos} \label{sec_aod}

We reproduced the critical dynamo numbers $R_\al$ further for the dipolar
$(l=1)$ mode of an isotropic $\al^2$-dynamo with $\al_{rr}=\al_{\th\th}=
\al_{\phi\phi}=R_\al\sin N\pi(r-r_i)$ and $N=1,2$ in a spherical shell of inner
and outer radius $r_i$ and $r_0$ with $r_0-r_i=1$ and $r_i/r_0=0.35$ and $0.8$
surrounded by a vacuum and either an insulating or a conducting inner core, as
reported in Table~2 of \citet{sz01}. With this test we treated in particular
two different aspect ratios of a spherical shell (a thick and a thin one) and
two different molecular diffusivities (insulating or conducting) of the inner
core.

Finally we applied our method to an $\al^2\Om$-dynamo of \citet{jw06} who
employ the classical eigenvalue treatment for the poloidal and toroidal scalars
$P$ and $T$ expanded in spherical harmonics in the angular coordinates and in
Chebychev polynomials in $r$-direction. We set $u_r=u_\th=0$, $u_\phi=R_\Om
r^3\sin^3\th$, $\al_{ij}=\beta_{ijk}=0$, except
$\al_{rr}=\al_{\th\th}=\al_{\phi\phi}=R_\al \sin2\pi(r-r_i)/(r_0-r_i)\cos\th$
with $r_i=0.5$, $r_0=1$, embedded in a vacuum inside and outside, and
$R_\al=\al_0r_0/\eta=10$ and $R_\Om=u_0r_0/\eta=1000$. Some results obtained by
the new and the classical methods are compiled in Table~\ref{tab_jie}. Numbers
in parentheses $(\dots\,,\,\dots)$ are the real and imaginary parts of complex
eigenvalues. The real part denotes the growth rate, the imaginary part the
frequency of the mode in units of $\eta/r_0^2$. The modes are axisymmetric
$(m=0)$, the fundamental mode is monotonously growing and symmetric (indicated
by $S$) with respect to the equator, and the fourth overtone is damped,
oscillatory and antisymmetric (indicated by $A$). Modes with higher $m$ are
more strongly damped. $n_\mathrm{max}$ refers to the maximum radial number of
the decay modes (spherical Bessel functions) for the new method and to the
maximum degree of the Chebychev polynomials for the code of \citet{jw06},
respectively. Since the modes have smaller length scales in latitudinal than in
radial direction, higher values of $l$ than of $n$ are required for
convergence. We find remarkably similar convergence of the eigenvalues for both
methods. This also applies to modes with higher $m$. Of course we have also
verified that the eigenfunctions obtained with the two methods are identical.

\section{Geodynamo models} \label{sec_gd}

Having proven that the new method works correctly and efficiently, we now apply
it to determine the eigensolutions of the dynamo operator with mean-field
coefficients obtained from self-consistent numerical simulations of the
geodynamo. For a recent review of numerical geodynamo simulations, see
\citet{cw07}.
\citet{srsrc07} developed an efficient method of calculating all tensorial
mean-field coefficients $\balpha$ and $\bbeta$ and compared the results of
mean-field and direct numerical simulations of the geodynamo.
We plan to use the eigenmodes of the dynamo equation to decompose the magnetic
field of the numerical simulations and to determine the statistical properties
of the mode coefficients \citep{h09} to analyse the working of the geodynamo.

\subsection{Benchmark dynamo} \label{sec_bd}

We examine a quasi-steady geodynamo model which has been used before as a
numerical benchmark dynamo \citep[case 1]{c01}. The governing parameters are
Ekman number $E=10^{-3}$, Rayleigh number $Ra=100$, Prandtl number $Pr=1$, and
magnetic Prandtl number $Pm=5$. The convection pattern is columnar with a
natural 4-fold azimuthal symmetry and is stationary except for an azimuthal
drift. The intensity of the fluid motion is characterised by a magnetic
Reynolds number of $R_m\simeq40$, defined with a characteristic flow velocity,
the thickness of the convecting shell, and the molecular magnetic diffusivity.
The magnetic energy density exceeds the kinetic one by a factor of 20.

\begin{figure}
\resizebox{\hsize}{!}{\includegraphics{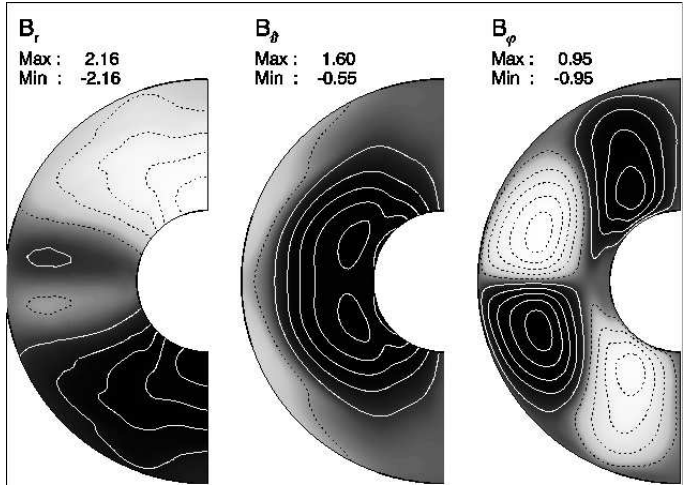}} \caption{Magnetic
field structure of the fundamental antisymmetric eigenmode for the benchmark
dynamo. Compare with Fig.~10 of \citet{srsrc07}. For each plot the grey scale
is separately adjusted to its maximum modulus with white as negative and black
as positive. The contour lines correspond to $\pm0.1$, $\pm0.3$, $\pm0.5$,
$\pm0.7$, and $\pm0.9$ of the maximum modulus.} \label{fig_bench1}
\end{figure}

In \citet{srsrc07}, the mean-field coefficients are derived from the numerical
simulation. We solved the dynamo equation with these mean-field coefficients by
the new method and obtained the eigenvalues and eigenfunctions. Since the
coefficients are spatially variable to a considerable degree, converged
solutions require high truncation levels in $n$ and $l$. The eigenvalues of the
first two modes are shown in Table~\ref{tab_bench1}. Beyond $n_\mathrm{max}
\simeq 20$ and $l_\mathrm{max} \simeq 20$, the eigensolution of the first mode
does not change significantly and is displayed in Fig.~\ref{fig_bench1}. The
convergence of the second mode requires a larger $l_\mathrm{max}$ of about
$32$. The results for high values of $n_\mathrm{max}$ may be affected by the
spatial variation of the mean-field coefficients and would require more than 66
radial quadrature points to compute the matrix elements.

A comparison of Fig.~\ref{fig_bench1} with its counterpart Fig.~10 of
\citet{srsrc07} shows that the field of the antisymmetric fundamental mode
resembles the field of an initial-value mean-field dynamo calculation
remarkably well as it does the axisymmetric component of the direct numerical
simulation. The mode here grows slightly with a rate around
$\lam_0\simeq4.2\eta/L^2$, the field of the initial value calculation decays
slightly with a rate of approximately $-0.25\eta/L^2$, while the solution of
the direct numerical simulation is stationary\footnote{In \citet{srsrc07} the
mean flow entered with a sign error into the initial value calculation, leading
to a stronger decay of $-3.5\eta/L^2$. We apologise and correct this value
here.}. Here $L=r_0-r_i=1$ is the thickness of the spherical shell. The
difference in these rates between the eigenvalue and initial value calculation
comes from the higher numerical diffusivity of the latter at the chosen
resolution of 33 radial and 80 latitudinal grid points. The difference is
actually small, much less than one effective decay rate, because the relevant
turbulent diffusivity, described by the $\bbeta$ coefficient with values up to
$33\eta$, is much higher than the molecular one.

\begin{table}
\caption{Eigenvalues in units of $\eta/L^2$ of the first two eigenmodes of the
benchmark dynamo.} \label{tab_bench1}
\begin{center}
\begin{tabular}{cccccc}
\hline\hline
$n_\mathrm{max}$ & $l_\mathrm{max}$ & $\lam_0 \; (A)$ & $\lam_1 \; (S)$ \\
\hline
12 & 12 & $+$4.960 & $-$8.605 \\
16 & 16 & $+$4.235 & $-$8.110 \\
20 & 20 & {\bfseries $+$4.180} & $-$8.362 \\
24 & 24 & $+$4.195 & $-$6.620 \\
28 & 28 & $+$4.255 & $-$7.275 \\
32 & 32 & $+$4.275 & $-$6.015 \\
\hline
16 & 32 & $+$4.382 & $-$6.777 \\
\hline
\end{tabular}
\end{center}
\tablefoot{The symmetry with respect to the equator is marked $A$ for
antisymmetric and $S$ for symmetric. The eigensolution marked bold is displayed
in Fig.~\ref{fig_bench1}.}
\end{table}

Besides the true physical eigenmodes, we find growing unphysical spurious
eigenmodes. Their eigenvalues depend strongly on the resolution, and their
eigenfunctions are highly structured. We attribute their appearance to a
locally confined inappropriate parametrisation of the mean electromotive force
by the mean-field coefficients $\balpha$ and $\bbeta$ \citep{srsrc07}. The
spurious modes are present neither in the initial value calculation nor in the
following example of a time-dependent dynamo, because of a higher numerical and
molecular diffusivity, respectively.

\subsection{A time-dependent dynamo in the columnar regime} \label{sec_tdd}

The next example has stronger forcing with parameters $E=10^{-4}$, $Ra=334$,
$Pr=1$, and $Pm=2$. The numerical simulation by \citet[case 2]{ocg99} shows a
highly time-dependent, but still dominantly columnar convection characterised
by a magnetic Reynolds number of $R_m\simeq88$. The magnetic energy exceeds the
kinetic energy by a factor of three. The magnetic field has a strong axial
dipole contribution. Although chaotically time-dependent, the velocity field is
symmetric and the magnetic field antisymmetric with respect to the equatorial
plane.

\begin{table}
\caption{Eigenvalues in units of $\eta/L^2$ of the first two antisymmetric
eigenmodes of the temporally averaged dynamo operator obtained from the
time-dependent dynamo (case 2, Sect.~\ref{sec_tdd}).} \label{tab_case2}
\begin{center}
\begin{tabular}{ccccc}
\hline\hline
$n_\mathrm{max}$ & $l_\mathrm{max}$ & $\lam_0$ & $\lam_3$ \\
\hline
12 & 12 & ($-$4.520\,,\,0.0) & ($-$35.046\,,\,$\pm$10.118) \\
16 & 16 & ($-$4.278\,,\,0.0) & ($-$35.090\,,\,$\pm$10.256) \\
20 & 20 & ($-$4.112\,,\,0.0) & ($-$34.578\,,\,$\pm$10.096) \\
24 & 24 & ($-$3.930\,,\,0.0) & ($-$34.770\,,\,$\pm$9.874) \\
28 & 28 & ($-$3.880\,,\,0.0) & ($-$34.766\,,\,$\pm$10.236) \\
32 & 32 & ($-$3.868\,,\,0.0) & ($-$34.804\,,\,$\pm$10.318) \\
\hline
16 & 32 & {\bfseries ($-$3.874\,,\,0.0)} & {\bfseries ($-$34.830\,,\,$\pm$10.310)} \\
\hline
\end{tabular}
\end{center}
\tablefoot{The eigensolutions marked bold are displayed in
Fig.~\ref{fig_case2}.}
\end{table}

\begin{figure}
\resizebox{\hsize}{!}{\includegraphics{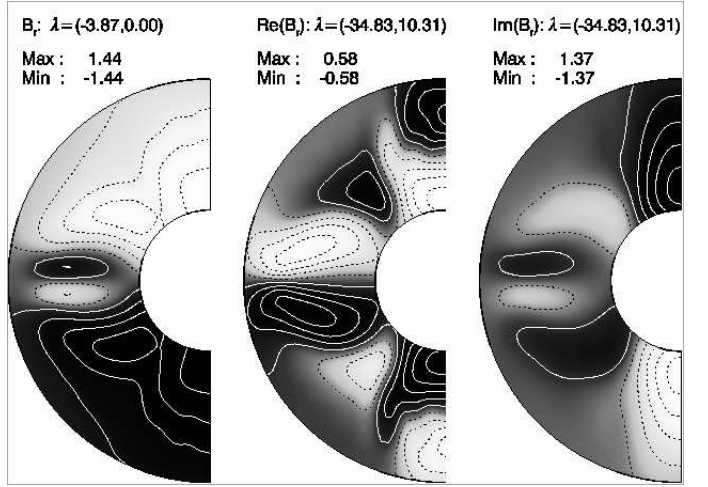}} \caption{Radial
component of the first two antisymmetric eigenmodes for the case2 dynamo. Left:
fundamental mode; middle: real part of the first overtone; right: imaginary
part of the first overtone. Grey scales and contours as in
Fig.~\ref{fig_bench1}.} \label{fig_case2}
\end{figure}

The mean-field coefficients are obtained as before by the test-field method of
\citet{srsrc07}. The coefficients are now of course also highly time-dependent.
A time average yields coefficients that roughly resemble those for the
benchmark dynamo, although there are differences in some profiles and
amplitudes.

For the time-averaged dynamo operator the eigenvalues of the first two
antisymmetric eigenmodes for various values of $n_\mathrm{max}$ and
$l_\mathrm{max}$ are shown in Table~\ref{tab_case2}. It seems that a value of
$n_\mathrm{max}\simeq16$ is sufficient for convergence, while
$l_\mathrm{max}\simeq32$ is needed. Figure~\ref{fig_case2} shows the
eigenfunctions of these modes. The eigensolutions for $\lam_1=(-6.298\,,\,0.0)$
and $\lam_2= (-28.712\,\,,\pm5.364)$, values for $n_\mathrm{max}=16$ and
$l_\mathrm{max}=32$, are symmetric with respect to the equator.

An initial-value, mean-field dynamo calculation with the same mean velocity and
dynamo coefficients shows a slighly decaying solution with a decay rate of
approximately $-5.9\eta/L^2$ which is to be compared with the eigenvalue
$\lam_0\simeq-3.87\eta/L^2$ of the fundamental mode. Again, the turbulent
diffusivity exceeds the molecular one by a factor of up to 23 in this case. The
difference in the decay rates is therefore much less than one effective decay
rate. As for the benchmark dynamo, the profile of the antisymmetric fundamental
mode is again remarkably similar to the solution of the initial value
calculation and to the axisymmetric component of the direct numerical
simulation.

A decomposition of the actual magnetic field of the simulation by \citet[case
2]{ocg99} in eigenfunctions of the time-averaged dynamo operator, i.e.,
$\bB(\br,t)=\sum_i c_i(t)\bB_i(\br)$, shows that the antisymmetric fundamental
mode contributes to about 75 percent and, together with the first antisymmetric
overtone (see Table~\ref{tab_case2} and Fig.~\ref{fig_case2}), to about 85
percent of the total magnetic energy. The variability in time of the magnetic
field of the direct numerical simulation is reflected in the variability of the
expansion coefficients. More details are presented in \citet{ssch10}.

\section{Conclusions and outlook} \label{sec_concl}

We presented a new method for computing the eigenvalues and eigenfunctions of
the induction and the dynamo equation. The method is based on the
biorthogonality of the adjoint electric current and the vector potential with
an inner product defined by a volume integral over the fluid domain. The
advantage of the method is that the velocity and dynamo coefficients do not
have to be differentiated. The method is therefore well-suited for spatially
strongly variable dynamo coefficients. 

We tested the new method against the classical treatment and proved that it
works correctly and efficiently. We applied it to two cases with dynamo
coefficients derived from direct numerical simulations of the geodynamo. The
obtained dynamo eigenmodes are promising candidates for decomposing the
magnetic field of the numerical simulations and for analysing the statistical
properties of the mode coefficients as proposed by \citet{h09}.

\begin{acknowledgements}
We thank Ulrich Christensen, Johannes Wicht, and Robert Cameron for many useful
discussions and support. We further thank the referee, Matthias Rheinhardt, for
his detailed comments that helped to improve the paper.
\end{acknowledgements}

\appendix

\section{Free magnetic decay modes in a sphere or spherical shell
embedded in vacuum} \label{dm}

We decompose the magnetic field in its poloidal and toroidal components
\be \bB=\bn\t\bn\t P\hr+\bn\t T\hr \ee
with defining scalars $P(r,\th,\phi)$ and $T(r,\th,\phi)$ and unit vector in
radial direction $\hr=(1,0,0)$ in spherical coordinates $(r,\th,\phi)$. The
equation of free magnetic decay
\be \frac{\p\bB}{\p t}=-\eta\bn\t\bn\t\bB \ee
with constant magnetic diffusivity $\eta$ then reads
\be \frac{\p}{\p t}(P,T)=\eta\left(\Delta_H+\frac{\p^2}{\p r^2}\right)(P,T) \ee
where $\Delta_H$ is the horizontal Laplacian
\be \Delta_H=\frac{1}{r^2\sin^2\th}\,\frac{\p}{\p\th}\sin\th\,\frac{\p}{\p\th}
+\frac{1}{r^2\sin^2\th}\,\frac{\p^2}{\p\phi^2} \;. \ee
The solutions are the free magnetic decay modes
\bea P_{lmn} &=& \exp\left(\lam_{ln}^Pt\right)xf_l(p_{ln}x)\,Y_l^m(\th,\phi) \;, \\
T_{lmn} &=& \exp\left(\lam_{ln}^Tt\right)xg_l(t_{ln}x)\,Y_l^m(\th,\phi) \eea
with $x=r/r_0$ where $r_0$ is the radius of the sphere. The growth rates are
given by
\be \lam_{ln}^P=-\eta p_{ln}^2/r_0^2 \;\;\mathrm{and}\;\; \lam_{ln}^T=-\eta t_{ln}^2/r_0^2 \ee
and are independent of the azimuthal degree $m$. The constants $p_{ln}$ and
$t_{ln}$ are
\be p_{ln}=j_{l-1,n} \;\; \mathrm{and} \;\; t_{ln}=j_{l,n} \ee
where $j_{l,n}$ is the $n$-th zero of $j_l$. The $Y_l^m$ are the spherical
harmonics and normalised to unity by taking
\be Y_l^m(\th,\phi)=\left(\frac{4\pi}{2l+1}\frac{(l+m)!}{(l-m)!}\right)^{-1/2}
P_l^m(\cos\th)\,e^{im\phi} \ee
using Ferrer's definition of the Legendre functions of first kind $P_l^m$ with
degree $l$ and order $m$.

For a sphere embedded in vacuum the radial functions are given by
\bea f_l(p_{ln}x) &=& \left\{ \begin{array}{ll} a_{ln}\,j_l(p_{ln}x) & 0\le x\le 1 \\[0.5mm]
a_{ln}\,j_l(p_{ln})x^{-l} & x\ge 1 \end{array} \right. \\
g_l(t_{ln}x) &=& \left\{ \begin{array}{ll} b_{ln}\,j_l(t_{ln}x) & 0\le x\le 1
\\[0.5mm] 0 & x\ge 1 \end{array} \right. \eea
with the spherical Bessel functions of first kind $j_l$. This ensures regularity in the
origin of the sphere, vanishing toroidal component at its outer boundary and
smooth transition of the poloidal component to a potential field in the vacuum
outside.

For a spherical shell with inner radius $r_i$ $(x_i=r_i/r_0)$ and outer radius
$r_0$ $(x_0=1)$ embedded in vacuum the radial functions inside the shell are
given by
\be f_l(p_{ln}x)=j_l(p_{ln}x)-y_l(p_{ln}x)j_{l+1}(p_{ln}x_i)/y_{l+1}(p_{ln}x_i) \ee
and
\be g_l(t_{ln}x)=j_l(t_{ln}x)-y_l(t_{ln}x)j_l(t_{ln}x_i)/y_l(t_{ln}x_i) \;, \ee
and the constants in the arguments are the roots of
\be j_{l+1}(p_{ln}x_i)y_{l-1}(p_{ln})-j_{l-1}(p_{ln})y_{l+1}(p_{ln}x_i)=0 \ee
for $p_{ln}$ and of
\be j_l(t_{ln})y_l(t_{ln}x_i)-j_l(t_{ln}x_i)y_l(t_{ln})=0 \ee
for $t_{ln}$. Here $y_l$ are the spherical Bessel functions of second kind.

The magnetic field of the decay modes $\bB_i$ is obtained by inserting the
spatial parts of the defining scalars $P_{lmn}$ and $T_{lmn}$, respectively,
into (A.1). Here we have comprised the three indices into one. The decay modes
are self-adjoint on $V+E$, so that the adjoint functions are obtained simply by
complex conjugation: $\hB_k=\bB_k^*$ and likewise $\hJ_k=\bJ_k^*$.
Normalisation on $V+E$, i.e.,
$(\bB_k^*,\bB_i)_{V+E}=(\bJ_k^*,\bA_i)_V=\delta_{ki}$, is thus straightforward.
For a unit sphere the radial functions are normalised to unity by scaling the
$f_l$ with
\be a_{ln}=\left(\frac{1}{2r_0}l(l+1)j_{l-1,n}^2\,j_l^2(j_{l-1,n})\right)^{-1/2}\ee
and the $g_l$ with
\be b_{ln}=\left(\frac{r_0}{2}l(l+1)j_{l+1}^2(j_{l,n})\right)^{-1/2} \;. \ee
For a spherical shell the normalisation constants are more lengthy expressions,
which we suppress here.

The free magnetic decay modes form a complete and orthogonal set of functions,
and they obey the boundary conditions of the magnetic field between the dynamo
volume $V$ and the exterior vacuum $E$.

We mention for completeness that the poloidal decay modes are not self-adjoint
on $V$, i.e., $(\bB_k^*,\bB_i)_V \ne \delta_{ki}$. If we like to work with an
inner product defined on $V$, the adjoint functions $\hB_k$ can be constructed
by requiring $(\hB_k,\bB_i)_V=\delta_{ki}$, similar to the one described in
Sect.~\ref{ae}.

\end{document}